\documentclass[twocolumn,showpacs,prl]{revtex4}
\usepackage{amssymb}
\usepackage{graphicx}
\usepackage{dcolumn}
\usepackage{bm}
\usepackage{amsmath}

\begin{document}

%\twocolumn[
%\hsize\textwidth\columnwidth\hsize\csname
%@twocolumnfalse\endcsname

\title{Bragg spectroscopy of the multi-branch Bogoliubov spectrum \\
of elongated Bose-Einstein condensates}
\author{J. Steinhauer$^1$, N. Katz$^1$, R. Ozeri$^1$, N. Davidson$^1$,
C. Tozzo$^2$, and F. Dalfovo$^{2,3}$}
\address{$^1$ Department of Physics of Complex Systems,\\
Weizmann Institute of Science, Rehovot 76100, Israel}
\address{$^2$ Dipartimento di Matematica e Fisica, \\
Universit\`a Cattolica del Sacro Cuore, via Musei  41, Brescia, Italy}
\address{$^3$ INFM Unit\`a di Brescia and BEC-INFM Trento, Italy}

%\maketitle

\begin{abstract}
We measure the response of an elongated Bose-Einstein condensate
to a two-photon Bragg pulse. If the duration of the pulse is long,
the total momentum transferred to the condensate exhibits a
nontrivial behavior which reflects the structure of the underlying
Bogoliubov spectrum. It is thus possible to perform a
spectroscopic analysis in which axial phonons with different
number of radial nodes are resolved. The local density
approximation is shown to fail in this regime, while the observed
data agrees well with the results of simulations based on the
numerical solution of the Gross-Pitaevskii equation.
\\
\end{abstract}

\maketitle
% ]

The crossover between phonon and single-particle excitations in
the Bogoliubov spectrum of a weakly interacting Bose gas
\cite{bogoliubov} is one of the main ``textbook" concepts that can
be directly tested in the case of trapped Bose-Einstein condensed
gases. Bogoliubov quasiparticles have been already produced in
elongated condensates by using two-photon Bragg scattering
\cite{stamperkurn,vogels,steinhauer,ozeri}. For excitations with
frequency $\omega$ and wavevector $k$, the phononic character has
been checked by observing that the static structure factor $S(k)$
is less than $1$ \cite{stamperkurn,steinhauer}, by measuring the
quasiparticle amplitudes $u_k$ and $v_k$ \cite{vogels,brunello1},
and by showing that the dispersion relation $\omega(k)$ is linear
at low $k$ \cite{steinhauer,ozeri}.

In all of these cases, the local density approximation (LDA) has
been used to adapt the Bogoliubov theory of uniform gases to the
actual inhomogeneous condensates. This approach is expected to be
accurate for large condensates, where the density profile varies
smoothly on the scale of the excitation wavelength and the system
behaves locally as a piece of uniform gas with a local Bogoliubov
spectrum \cite{zambelli,brunello2}. This applies, for instance, to
axial excitations of elongated condensates, when the wavelength of
the excited states is much smaller than the size of the system
along the major axis. These excitations can then be classified
with a continuous wavevector $k$. However, the finite transverse
size of the condensate also produces a discreteness of the
spectrum, which is ignored in LDA (see \cite{penckwitt} for a
recent classification of normal modes in anisotropic condensates
and \cite{fedichev,zaremba} for the limiting case of an infinite
cylinder).

In this Letter we show that the response of
the condensate to a Bragg pulse is indeed significantly affected by
the radial degrees of freedom. In particular, if the duration of the
pulse is longer than the radial trapping period, the condensate
responds resonantly at the frequencies $\omega_{n_r}(k)$ of axial
quasiparticles with $n_r$ nodes in the radial direction. By using
Bragg pulses longer than in previous measurements
\cite{stamperkurn,steinhauer}, we resolve this multi-branch spectrum,
finding good agreement with the predictions of Gross-Pitaevskii (GP)
theory.

As described in Ref. \cite{steinhauer}, our condensate consists of
$N=10^5$ atoms of $^{87}$Rb, with a thermal fraction of 5\% or
less.  The radial and axial trapping frequencies are $\omega_\perp
= 2\pi (220$Hz$)$ and $\omega_z= 2\pi (25$Hz$)$ respectively. The
radial and axial Thomas-Fermi radii of the condensate are $R=
3.1\mu$m and $Z=27.1 \mu$m.

We excite the condensate by using two Bragg beams with
approximately parallel polarization, separated by an angle
$\theta$.  The Bragg beams illuminate the entire condensate for a
time $t_{B}$. \ The beams have a frequency difference $\omega$
determined by two acousto-optic modulators. \ If a photon is
absorbed from the higher-frequency beam and emitted into the
other, an excitation is produced with energy $\hbar \omega $ and
momentum $\hbar {\bf k}$, where $k=2k_{p}\sin \left( \theta
/2\right)$, and $k_{p}$ is the photon wave number. The wave vector
${\bf k}$ is adjusted to be along the $z$-axis. To measure a
single point on the excitation spectrum $\omega (k)$, $k$ is fixed
by $\theta $, and $\omega$ is varied.

The measured quantity is the total momentum transferred to the
condensate along $z$. This is obtained by switching off the
trapping potential after the Bragg pulse and taking absorption
images of the density distribution of the expanding condensate. In
Fig. 1 we show typical results for short Bragg pulses of duration
$t_B =1$ msec. The observed momentum  $P_z$ is plotted as a function
of $\omega$ for three different values of $k$. In Ref.~\cite{steinhauer}
we made Gaussian fits to these type of curves,
taking the position of the maximum to be the quasiparticle
dispersion $\omega(k)$ and showing that this dispersion was
consistent with the Bogoliubov spectrum in the LDA.
\begin{figure}[h]
\begin{center}
\includegraphics[width=3.3in]{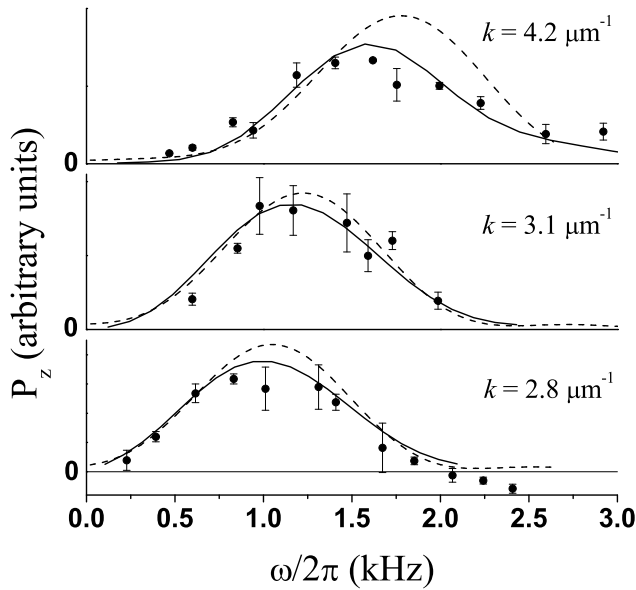}
\end{center}
\caption{$P_z$, in arbitrary units, as a function of $\omega$ for
short pulses ($t_B =1$ ms). Bottom, center, and top plots
correspond to different $k$-values. Points with error bars are the
measured values. Solid lines are the results of GP simulations.
The dashed lines are the LDA predictions.}
\end{figure}

A deeper analysis of the same results can be performed by numerically
solving the GP equation for the order parameter of the condensate
$\Phi({\bf r},t)$ \cite{review}:
\begin{equation}
i \hbar \partial_t \Phi = \left( - { \hbar^2 \nabla^2 \over 2m } +
V + g |\Phi|^2 \right) \Phi
\label{eq:TDGP}
\end{equation}
where $g=4\pi \hbar^2 a/m$ is determined by the $s$-wave scattering
length $a$. The external potential can be taken as the sum of the
harmonic confinement and the Bragg potential,
\begin{equation}
V({\bf r},t) = {m\over 2} ( \omega_\perp^2 r_\perp^2 + \omega_z^2
z^2) + \theta(t) V_B \cos(kz-\omega t)
\label{eq:Vext}
\end{equation}
where $r_\perp^2=x^2+y^2$ and $\theta(t)$ is equal to $1$ in the
interval $0 < t < t_B$ and $0$ outside.

The ground state at $t=0$ can be found as the stationary solution
of Eq.~(\ref{eq:TDGP}).  Then, the time dependent GP equation can
be solved at $t>0$ to simulate the Bragg process
\cite{brunello2,blackie}. We take advantage of the axial symmetry
of $V$ to map the order parameter into a two-dimensional grid of
points $N_\perp \times N_z$ (typically, $64 \times 1024$) and
evolve it by means of a Crank-Nicholson differencing method with
alternating direction implicit algorithm, as in \cite{modugno}.

The momentum transferred to the condensate can be calculated from
$\Phi$ through the definition $P_z = (- i \hbar/2) \int d{\bf r}
\, \Phi^* \partial_z \Phi + {\rm c.c.}$. Fig.~1 shows good
agreement between the results of the GP simulations (solid lines)
and the experimental data, for short pulses.  The strength $V_B$
in the GP equation is used as a free parameter, so the comparison
is restricted to the position and shape of the peak.  In
principle, $V_B$ could be determined by using the experimental
parameters for the power, area, and direction of polarization of
the two laser beams, but this estimate might be affected by a
significant uncertainty.

The LDA curves (dashed lines in Fig. 1), which are obtained by
using the local Bogoliubov spectrum to calculate $P_z(k,\omega)$
in the linear response regime as in \cite{brunello2}, are also
reasonably close to the GP predictions. For shorter $t_B$ we
checked that the accuracy of the LDA is even better, as discussed
in \cite{brunello2}.

A rather different situation is found for longer pulses, with
$t_B$ of the order of the radial trapping period (about $4.5$
msec) \cite{note2}.  Fig.~2 shows the momentum transferred in the
GP simulations (solid curves) for various time durations $t_B$.
The lowest curve, after $1$ msec, is a broad peak as in Fig.~1,
close to the LDA (dashed curves). For later times however, a
multi-peak structure appears, strongly deviating from LDA. We
find a similar behavior for all values of $k$ in our simulations,
and for a wide range of intensities $V_B$.
\begin{figure}[h]
\begin{center}
\includegraphics[width=3.3in]{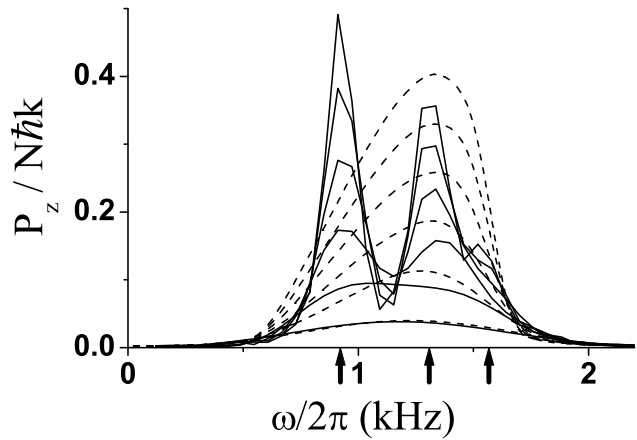}
\end{center}
\caption{The time dependence of the momentum transferred during
the Bragg pulse, for $k=3.1 \mu m^{-1}$. Solid curves are the
results of GP simulations with $V_B=0.2 \hbar \omega_\perp$
plotted at various times.  Starting with the lowest solid curve,
$t_B = 1, 2, 3, 4, 5$ and $6$ msec. Dashed curves are the
corresponding LDA predictions. The arrows indicate the GP
predictions for the frequencies of the normal modes of the
condensate, having $0$, $1$ and $2$ radial nodes and the same
axial wave vector $k$. }
\end{figure}

Figures 3a and 3b show measurements of $P_z$ for $k=1.4 \mu
m^{-1}$ and $3.1 \mu m^{-1}$ respectively, with long pulses of
duration $10$ msec and $6$ msec, respectively.  The intensities of
the Bragg beams are adjusted so that the number of excitations
created on resonance is no more than roughly 25\% of the number of
atoms in the condensate. Each point in Fig.~3 is an average of
about 5 measurements. The results of GP simulations are indicated
by the upper solid curves, which are obtained by choosing $V_B=0.2
\hbar \omega_\perp$ and $0.35 \hbar \omega_\perp$ for Figs. 3a and
3b, respectively.  Both the measurements and the GP simulations
show multiple peaks, due to the excitations of quasiparticles
with different $n_r$.

In Fig.~3, we see good agreement between the locations of the
peaks in the experimental and GP results, but the experimental
peaks are broader due to noise. The major source of noise is the
sloshing of the condensate in the trap. Specifically, sloshing of
speed $v_{sl}$ in the axial direction yields a Doppler shift in
the light potential of frequency $kv_{sl}/2\pi$. This gives a
frequency noise that significantly broadens the peaks of $P_z$.
However, we can reduce this noise by estimating the velocity of
the condensate in the trap at the end of the Bragg pulse from its
position in the time-of-flight image, and by explicitly adding the
corresponding Doppler shift to the applied $\omega$ in the
laboratory frame. This correction requires that the Bragg pulse be
much shorter than the axial trap period ($40$ msec). This
requirement is sufficiently met by the 6 msec pulses employed for
$k=3.1 \mu m^{-1}$. Therefore, the data points of Fig.~3b contain
the correction.

We verify that the multi-peak structure reflects the fundamental
normal modes composing the multi-branch spectrum, and does not
depend significantly on the intensity $V_B$.  The lower curves of
Figs.~3a and 3b correspond to the same simulations as the upper
curves, but with $V_B$ reduced by a factor $10$. In this case, the
system is excited in the linear response regime. The momentum
transferred is two order of magnitude smaller, but the resulting
curves have peaks at the same locations as for larger $V_B$.
Furthermore in the case of $k=3.1 \mu m^{-1}$ we repeat the
measurement and simulation of Fig.~3b with an intensity $30\%$
greater and find peaks again at the same location. This means that
nonlinear effects are not crucial in our observations.
Nevertheless, they might be interesting. Indeed, looking at
Fig.~3a one notices small differences between the lower and upper
curves, for small and large $V_B$ respectively.  The curve for
large $V_B$ displays two side peaks around the main peak at $0.4$
kHz. These side peaks, whose shape significantly depends on $t_B$,
are not visible in the linear regime of the small $V_B$ curve and
might be due to nonlinear effects.  Similar effects were found by
a one-dimensional simulation in Ref. \cite{band}.

\begin{figure}[h]
\begin{center}
\includegraphics[width=3.3in]{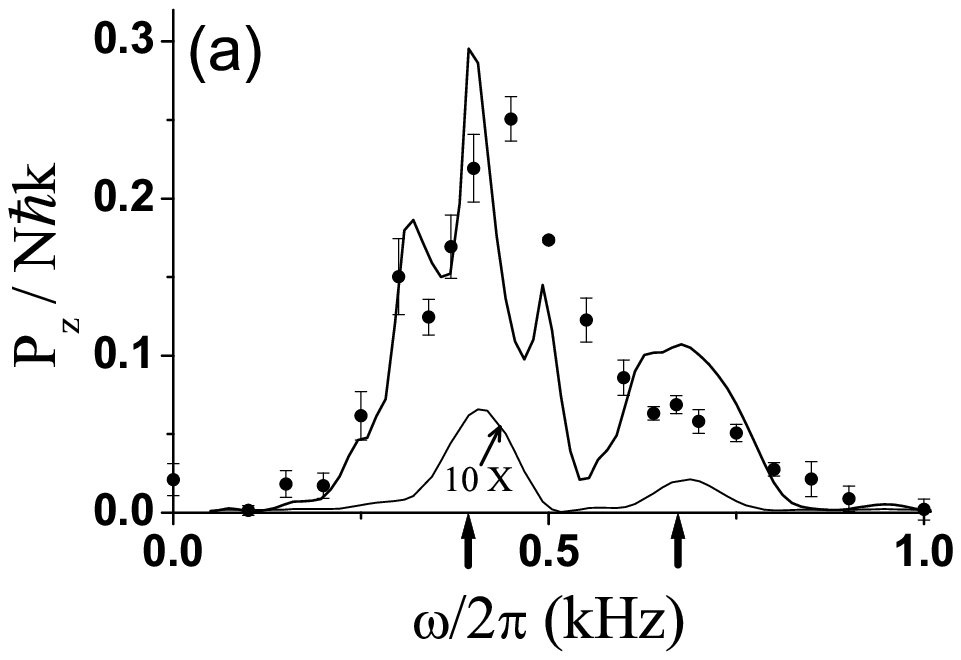}
\end{center}
\end{figure}
\begin{figure}[h]
\begin{center}
\includegraphics[width=3.3in]{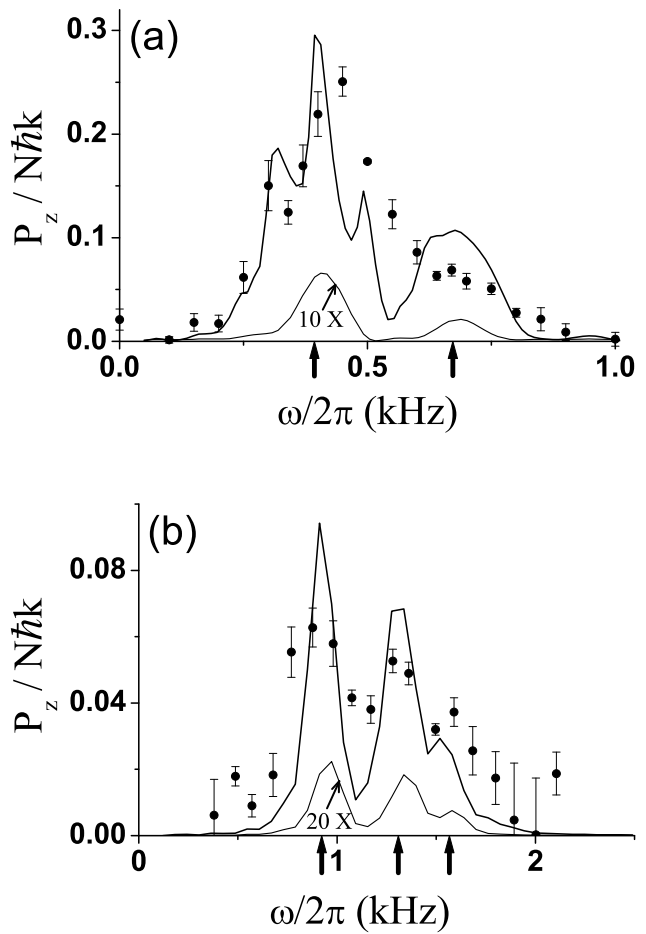}
\end{center}
\caption{$P_z$ in units of $N\hbar k$, as a function of $\omega$
for long Bragg pulses.  For (a) and (b), $t_B=10$ msec and 6 msec
respectively, and $k=1.4 \mu m^{-1}$ and $3.1 \mu m^{-1}$
respectively.  The circles are the measured values.  The upper
curve is the result of GP simulations with $V_B=0.35 \hbar
\omega_\perp$ and $0.2 \hbar \omega_\perp$, for (a) and (b)
respectively. The lower curve is the same but with $V_B$ ten times
smaller. For the lower curve, $P_z$ is multiplied by $10$ and $20$
for (a) and (b) respectively. The arrows on the $\omega$ axes
indicate the normal mode frequencies, as in Fig.~2.}
\end{figure}

Now we use the GP equation to directly find the multi-branch
Bogoliubov spectrum of the condensate. Specifically, we analyze
the oscillations in the condensate density induced by the Bragg
process. In the GP simulation we let the condensate freely
oscillate in the trap after the Bragg pulse and we perform a
Fourier analysis of the density variations. This analysis shows
that for each $k$ and $\omega$, the density oscillates as a
superposition of modes of frequency $\omega_{n_r}(k)$, which are
excited by the Bragg potential in Eq.~(\ref{eq:Vext}) due to the
inhomogeneity of the condensate in the radial direction. For
symmetry reasons, only modes with azimuthal angular momentum $m=0$
are excited. The calculated frequencies are shown as open circles
in Fig.~4 \cite{note3}.  The lowest branch corresponds to
Bogoliubov axial modes with no radial nodes. The second branch has
one radial node; it starts at $2 \omega_\perp$ for $k=0$, where it
corresponds to a purely radial breathing mode. The third branch
has two nodes \cite{note4}.

In the limit of small oscillations, all these states coincide with
the solutions of the linearized GP equation, i.e., the
generalization of the Bogoliubov equations to inhomogeneous
condensates. For an infinite cylinder these solutions were
calculated in Ref.~\cite{fedichev} and, in the hydrodynamic limit
($k$ much smaller than the inverse healing length $\xi^{-1}$), in
Ref. \cite{zaremba}. For $k \ll \xi^{-1}$, our multi-branch
spectrum turns out to be very close to the spectrum predicted in
Ref. \cite{zaremba}, indicated by dashed lines in Fig. 4 (also see
Fig. 1 of Ref. \cite{zaremba}). $\xi^{-1}$ is indicated by a
dotted line in Fig. 4.

\begin{figure}[h]
\begin{center}
\includegraphics[width=3.3in]{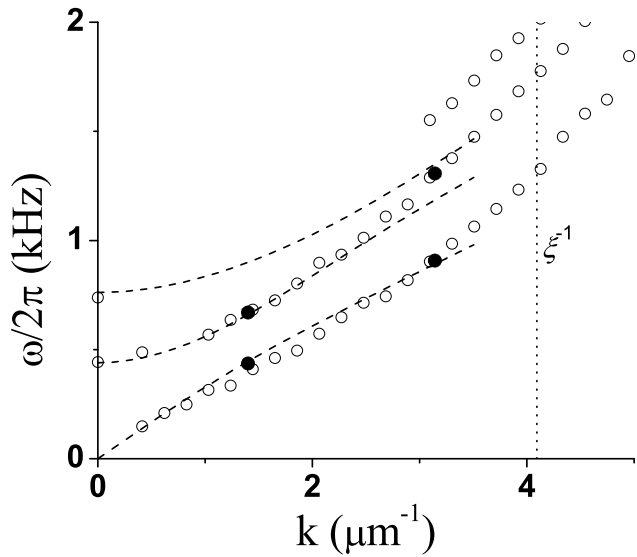}
\end{center}
\caption{Excitation frequencies vs. $k$ for a trapped BEC. The
open circles are the frequencies of the  normal modes with
$n_r=0,1,2,\dots$ nodes in the radial direction, as calculated
from the GP equation.  The filled circles are the position of the
maxima of the main peaks in $P_z$ for the long-pulse measurements
of Fig.~3.  The dashed lines are the prediction for an infinite
cylinder with $k \ll \xi^{-1}$. The dotted line indicates the
inverse healing length.}
\end{figure}

The calculated normal mode frequencies are shown as arrows on the
$\omega$ axes in Figs.~2 and 3. Both the theoretical curves and
the experimental results clearly show that the momentum is
resonantly transferred to the condensate when the Bragg frequency
is close to the Bogoliubov branches.

The locations of the measured peaks of Fig.~3 are also shown as
filled circles in Fig.~4. The good agreement between these
measured peaks and the simulated spectrum (open circles)
identifies the peaks in the observed spectrum as being axial
quasiparticles with $n_r$ radial nodes.  Bragg pulses can indeed
be used to perform high resolution spectroscopy of the
multi-branch Bogoliubov spectrum of trapped condensates in a
regime where the LDA is not applicable.

In conclusion, we have used long Bragg pulses to spectroscopically
measure the first and zeroth order radial modes in the phonon
spectrum of a BEC.  These high-resolution measurements agree well
with simulations of the Gross-Pitaevskii equation.  The local
density approximation fails to reproduce these multiple radial
modes.

The technique of long Bragg pulses could also be used to resolve
inherent dissipation mechanisms, such as inhomogeneous broadening
and coupling between modes, including Beliaev and Landau damping.
The technique could be particularly effective if the inhomogeneous
broadening could be overcome, by means of echo techniques, or the
use of non-quadratic optical dipole traps. Finally, very sensitive
GP simulations could be used in combination with long pulses, to
carefully study non-linear dynamics.

This work is partially supported by MIUR-COFIN2000 and the Israel
Science Foundation.


\begin{references}
\bibitem{note2} A longer timescale is fixed by the axial trapping
period ($40$ ms). For such durations other processes become
important (reflection of axial modes at the boundary,
center-of-mass motion, nonlinear effects, etc.).  This regime is
considered in \cite{blackie}.
\bibitem{note3} The calculated frequencies have an uncertainty
of the order of $0.05$ kHz, due to the finite time interval used
for the Fourier analysis.
\bibitem{note4} In this range of $k$ only three branches are visible
in the Fourier spectrum. The efficiency of the Bragg perturbation
(\protect\ref{eq:Vext}) in exciting higher branches increases with
$k$.
\end{references}
\end{document}